# SNA-BASED REASONING FOR MULTI-AGENT TEAM COMPOSITION

André Filipe de Moraes Batista[1] and Maria das Graças Bruno Marietto[2]

[1]Metropolitan-United Faculty (FMU), São Paulo, Brazil.
[2]Federal University of ABC (UFABC), São Paulo, Brazil.

## ABSTRACT

*The social network analysis (SNA), branch of complex systems can be used in the construction of multi-agent systems. This paper proposes a study of how social network analysis can assist in modeling multi-agent systems, while addressing similarities and differences between the two theories. We built a prototype of multi-agent systems for resolution of tasks through the formation of teams of agents that are formed on the basis of the social network established between agents. Agents make use of performance indicators to assess when should change their social network to maximize the participation in teams.*

## 1. Introduction

One of the research topics in the Multi-Agent Systems (MAS) area is the definition of models that represents social structures, such as organizations and alliances, in order to analyze more objectively the emergent behavior of open systems. Individuals who are related to each other by different types of relationships, such as dependencies on goals, conflicts over resources, similar beliefs and so on compose organizations and alliances.

Since the 80s several studies were carried out using the theory of MAS to represent models of social phenomena. Axelrod in [1] argues that the object of these studies is the use of the theory of MAS to break up simplistic definitions on a particular subject, due to the need of the model be mathematically tractable. Thus, social phenomena models normally used concepts such as homogeneity, ignoring interactions.

With the use of agent-based models, this research area has been benefited by being able to represent most of the behaviors of autonomous agents and the interactions between them. Agent-based models are in the most cases suitable for decentralized processing and decision making, particularly when individual interactions lead to the emergence of collective patterns, as well as complex systems. Therefore, there is a close relationship between MAS and Complex Systems Theory, and this relationship can be represented in the Social Network context.

An important question is how to represent these relationships, characterized by a high degree of dynamism. Dealing with social issues, MAS area was inspired mainly by the economic organizational theory and legal theory. Although, there is a lack of attention in the area of research that describes the relationships between individuals within human organizations and their dynamics. To this area, we give the name of Social Network Analysis (SNA). The social network analysis has emerged as a key technique in modern sociology, anthropology, social

International Journal of Artificial Intelligence & Applications (IJAIA) Vol. 6, No. 3, May 2015

psychology, communication studies, information science, organizational studies, and economics, among others.

This work aims to integrate the techniques of multi-agent systems and social network analysis, allowing an agent society make use of SNA metrics in the decision-making process considering attributes such as the position in the network and the value and the importance of an agent in the network. Since both techniques use the social element, the agent, the synergy between them will allow easy use of models based on social network analysis, so that the multi-agent systems become more reliable in the representation of social phenomena. For this, an application was developed representing process of selecting teams for solving tasks in a multi-agent society, in which agents have specific skills and their position in the social network will influence their performance in the multi-agent society.

This work is structured as follows: in Section presents a review of the issues of multi-agent systems and social network analysis area, while Section presents prospects of integration between both techniques. Section presents and discuss about the prototype of multi-agent system for a task-solving problem by agents teams making use of SNA technique. Section presents the final considerations of this work.

## 2. Background

In this section, we present a review of the literature on multi-agent systems (MAS) and social network analysis (SNA) research areas.

### 2.1. Multi-Agents Systems

Multi-agent systems consists in multiple agents that interact to each other in order to perform a specific set of tasks. The metaphor of intelligence used by these systems is the intelligent community in which social behavior is the basis for the system intelligence. From this, we can distribute the agents in areas of specialization and with the interaction among them a complex problem can be solved in a faster and more dynamically way[2].

The metaphor often used to structure the agents society is the human social groups, where expert teams can solve problems cooperatively, and the complexity exceeds the individual capacities of each of their members [3].

The autonomy of an agent allows it to take its own decisions to achieve its goals [4]. Thus, agents can get in and get out of the society, change their rules, roles, inter-dependent relationships with other agents, etc. This feature leads to a new generation of systems and distributed applications intrinsically dynamic, open and complex.

The notion of agents and multi-agent systems have been adopted in the modeling of various complex systems involving urban planning, biology, logistics and production, and many others [5, 6, 7, 8, 9, 10, 11, 12, 13]. Lynne and Nigel [14] propose a model based on agents for social networks. At the same step that the proposed model is simplistic, it can represent a wide variety of social networks. Teresa and Nina [15] implemented a Java tool that presents the dynamic model of social behavior associated with the recruitment of terrorists based on prescriptive models. They used concepts of MAS and SNA to represent this dynamic. Ronald et al. [16] proposed a model based on agents to analysis the social influence in the travel activities decision process. The agents have a travel schedule, and interact with each other in order to schedule social activities, in particular trade based on the nature of the activity, who will participate, time and

International Journal of Artificial Intelligence & Applications (IJAIA) Vol. 6, No. 3, May 2015

location. The structure formed between the agents described as a complex social network is the core of the decision process.

## 2.2. Social Network Analysis

Social Network Analysis (SNA) is a scientific research area derived from areas such as Sociology, Social Psychology and Anthropology. This area studies the relational links (a.k.a *relational tie*) between social actors. An actor in the SNA can be both individuals and companies, analyzed as individual units, or collective social units such as, for example, departments within an organization, public service agencies in a city, nation-states of the continent, among others. The SNA fundamentally differs from other studies in that the emphasis is not on the attributes (characteristics) of the actors, but the links between them.

Relations between actors pairs are made by relational ties or linkages. The most common types of links are: individual assessment (eg, friendship or respect); the transaction and the transfer of material resources (a purchase and sale transaction between two companies); the transfer of nonmaterial resources (the exchange of electronic messages) or not; the association or affiliation that occurs when actors participate in joint events (parties); interaction (sitting next to another person); the movement and the physical and social connection; links between formal roles (boss-subordinate authority loop in a company); biological relationships (father and son) and so on.

SNA is a method for enhancing the sharing of knowledge by analyzing the position and structure between actors, i.e their relationships. According to [17], the network analysis process considers basically two analytical perspectives that complement each other:

1. Egocentric - this type of analysis has attention in facing the particular node/actor (ego) and other nodes/actors in the network with others ego nodes that it maintains relations. Therefore, the number, the magnitude and the diversity of the direct or indirectly connections established with the ego define the other network nodes;
2. Socio centric - in this type of analysis, information about the pattern of links between all nodes/actors in the network is used broadly to identify reticulated subgroups with high degree of internal cohesion and the nodes that have similar positions on the network.

The interpretation of the results obtained with the SNA metric can be made from three points of view, namely:

1. Individual positions - from the viewpoint of the actors;
2. Whole network - from the viewpoint of the assembly of links forming the network;
3. clusters and components - from the viewpoint of the groups formed due to some kind of relationship.

The social network analysis metrics provides mathematical mechanisms to analyze a given society or group. Among the most significant metrics for understanding the role of an actor in the network, we can cite:

- Centrality Degree: it indicates the prestige and power that the author has in that particular community. In the SNA context, the power dimension is derived from the relational ties of an actor. The more relationship an actor makes, the more power it has.
- Betweenness Degree: it shows how much an actor is between others actors on the network. It helps to locate actors in positions of influence, those with the network information, etc. An actor with a high degree of betweenness may be functionally operating as a broker in the network;

- Closeness Centrality: shows the closeness of an actor to others on the network and thus its access to resources circulating in the network, based on the evaluation of the shortest path within the network.

Many of these metrics can be used in a multi-agent scenario, since it is composed of independent individuals who interact with each other. However, there is still a gap on the theory and practice on this issue, that is, how to implement and model a MAS with many different relationships between agents in order to capture and analyze them.

## 3. Social Network Analysis and Multi-Agent Systems

One of the biggest questions about the social network analysis is "it is known that social networks can be used to analyze human societies. But these can also be used for societies of agents?". An inconvenient and often questionable point, on social network analysis is that normally a transcription of the human social structures in sociological meanings is made from a clear and artificial manner, due to the complexities of individuals and their relationships. On the opposite side the simplicity in social terms of multi-agents systems suggests that the analysis of social networks can be applied to these, including may get even better results.

In the interconnection of two areas of study, one of the points of attention is the level of approach to be carried out. A multi-agent system is built from the specification of the smallest entity: the agent. The behavior and the structures are emerging consequences of their interactions. The same is true for Social Networks; the interaction of the actors will result in the emergence of the structure of the social network.

Many MAS consist of a complex network of autonomous and interdependent agents. In most of these systems, agents must select a set of other agents with whom will interact, based on factors such as limited resources, exchange of interests of knowledge, etc.

[18] Showed that the structure of existing artificial social network in multi-agent society, when used to govern the actions of the agents, is extremely connected with the performance of the multi-agent society.

Modeling scenarios in which agents are geared to social networks presents a set of challenges. First, agents must make adaptations in the network by decisions made based on local system information. This local information can give the agent a figure partially or completely wrong about the current environment. In this case, it is said that an agent has a limited horizon of the system. Since agents are organized as networks, and how various agents perform their local decisions (adapting your social network), these simultaneous changes may neglect the benefits of adaptation made by an agent.

Focusing on the social network adaptation strategy of an agent in a cooperative MAS (the agents cooperate towards a single goal), the following questions should be considered when modeling such agents, which are

- Local perception of overall performance: how an agent can estimate the collective performance of the organization? These estimates may be unreliable, since they are each based on partial views of the organization. A possible solution to this difficulty is to explore the communication skills of the agents. With use of some communication protocols, an agent can estimate whether its perception of overall performance is correct, or if it can use the perception of neighboring agents to improve it;

International Journal of Artificial Intelligence & Applications (IJAIA) Vol. 6, No. 3, May 2015

- Triggers of adaptation: when an agent must decide to adapt its structure of local connectivity? There are many possibilities of these triggers be fired. An agent, for example, may decide to adapt its structure based on estimates of performances in relation to the attendance of its goal;

- Network adaptation: how an agent decides which has to be removed connection and how it selects a new agent to establish a new social connection. A good strategy advocated by [18] is the adaptation based on reference where an agent asks for a reference about another agent for one of your neighbors.

## 4. Social Network Analysis' based Reasoning for Multi-Agent Team Composition

Based on [18], the proposed prototype on this work is a simple but intuitive multi-agent cooperative system. It is a multi-agent society for dynamic team formation of task resolvers.
This prototype presents a model of dynamic formation of teams, where teams of agents are formed spontaneously and in the decentralized form, as soon as the decision of an agent in adhering to a team is done by base in its local social network.

In the proposed model, a set of tasks are generated periodically and are globally advertised to the multi-agent society. The role of agents in this society is forming teams to solve these tasks. The participating agents of this society are involved for a social network. For an agent to be on a team, it must possess a social connection (that is, an edge in the network) with at least one member of the team.

Once this prototype focuses on the team formation process, when these teams are complete the task is considered done. Thus, in this model the multi-agent society consists of $N$ agents, $A = \{a_1, a_2, ..., a_N\}$, where each agent can be considered as a single node on the social network. The network is modeled as an adjacency matrix $E$, where each adjacency matrix elements $e_{ij}= 1$ if there is an edge between the agents $a_i$ and $a_j$, or $e_{ij}= 0$ otherwise. The degree (number of connections) of an agent $a_i$ is defined by

$$k_i = \sum_{a_j \in A} e_{ij} \qquad (1)$$

In multi-agent society, each agent has a single ability, $\sigma_i \in [1,\sigma]$, where $\sigma$ is the number of different types of abilities present in the society. During the process of forming teams, each agent can be in one of three states: uncommitted, committed, active. An agent is in the state uncommitted if it is available to join teams and therefore is not in any team. When an agent selects a task in which it is able to perform, this is in the committed state. When this agent effectively is accepted for the team, the same is in the active state.

Agents that are in the active state can no longer get out of the team until it is complete or a timeout is reached. In this case, the team was not formed in the required time to resolve the task. The tasks are advertised in multi-agent society in a fixed interval of time $\mu$. Each task is a vector $T_k$ of size $|k|$ with required skills. Each task is announced by a finite number of time $\gamma$, ensuring that resources (i.e., agents) linked to a task become available if the team is not formed.

The requirement of a team be an induced connected subgraph of the social network means that for any agent on the team, $a_i \in M_k$ - where $M_k$ is the $k$ team of this society, there should be at least some other agent $a_j \in M_k, i \neq j$ such that $e_{ij}= 1$. This implies that an agent in the uncommitted state is only eligible to enter in the committed state in two situations: 1) team startup, when the team is empty, and 2) joining a team, when at least one of the neighbors of the agent is in committed state related to this task.

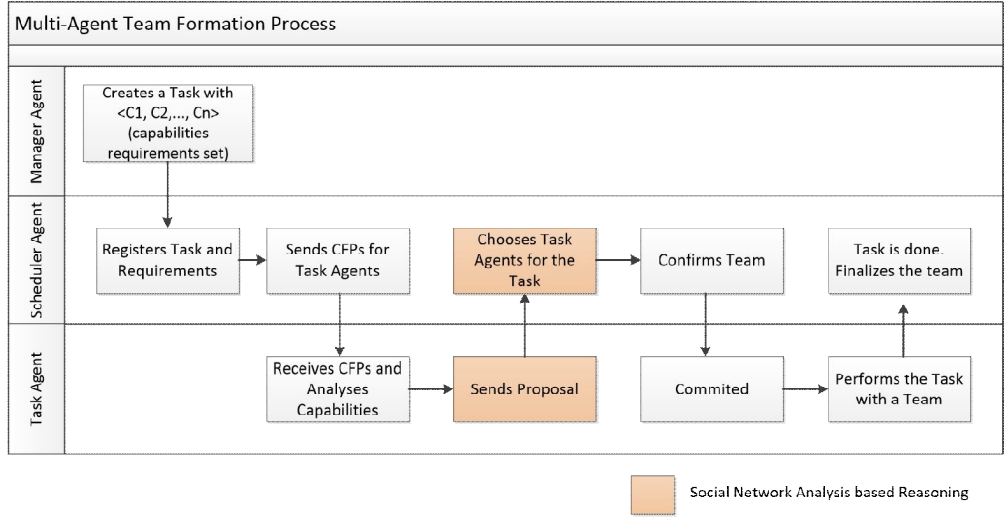

Figure 1: Process of forming teams of agents to task resolution, making use of social network metrics.

Unlike the model presented by cite [18], in this model all agents in the uncommitted state can try to belong to a team when they receive a task in which its ability is being requested. At this point, a Scheduler agent should coordinate the multi-agent society. It is up to this agent accepts or not the agents making use of SNA heuristics. Figure 1 presents an indicative flow chart of the multi-agent process, highlighting where the techniques and metrics from social network analysis are used.

In the flow shown in Figure 1 interactions, the role of the manager agent is generating tasks (skill set) and send them to the scheduler agent. The scheduler agent has the following set of functions: receive tasks sent by the manager agent and transmit them to all task agents; receive the proposals from task agents; select the agents to composition of the teams and check the conclusion (satisfactory or not) of the tasks.

Task agents receive tasks and if they have the necessary skills, send an application to the scheduler agent. Right now, they are awaiting a confirmation or a rejection message. If it is rejected then returns to receive tasks. If accepted, goes to the team and so it is waiting for the complete formation of the team.

Social strategies are implemented as follows:

1. Scheduler Agent: it knows the entire network of agents. When an agent task sends a message stating the interest in participating in the team, the scheduler agent checks whether the team is empty (in this case the agent is starting the team), or if the agent knows other agents that make up the team. This strategy that how the structure of social networks can be used for decision-making. Such approach can be used to solve problems of allocation in a general way. These problems have a high degree of complexity, and this occurs for two main reasons. First, find the optimal scheduling is a NP-hard problem. In addition, each scheduling problem has particular details, involving changes in the search algorithm of the

International Journal of Artificial Intelligence & Applications (IJAIA) Vol. 6, No. 3, May 2015

solutions. Thus, it becomes necessary to use some type of heuristic to reduce the search space. SNA metrics can be a valid heuristic in this situation;

2. Task Agent: each task agent has a local measure of performance. This measure will be the gear adaptation of the agent's social network strategy. This strategy is based on performance and reference [18]. The trigger of this adaptation is the measure of local performance $Y(a_i)$, which each agent $a_i$ has. This measure is the ratio between the number of times the agent participated effectively in a team (it was accepted for a team) by the number of attempts to participate. The measure of performance is considered valid when the agent tried to enter a team at least $v$ times. With each iteration, the agent $a_i$ chooses adapt their social network if this has a measure of performance valid, and if the measure of performance is below the average of the sum of the performance of all its neighbors. That is, an agent chooses adapt its social network if:

$$Y(a_i) < \frac{1}{k_i} \sum_{a_j \in A, e_{ij}=1} Y(a_j) \qquad (2)$$

If an agent decides to adapt its network based on the performance value, this adaptation is given both by performance and by reference. The agent will remove the connection to its immediate neighbor $a_j$ which has the smallest measurement of performance:

$$a_j = arg\ min_{a_m \in A, e_{im}=1}\ Y(a_m) \qquad (3)$$

The agent asks for a reference to the neighbor with the highest performance. Similarly, this agent will refer to its neighbor with the highest performance. Is $a_l$ the agent to whom the agent $a_i$ calls for reference, the agent will establish a new connection with $a_k$, the neighbor of greater performance of $a_l$:

$$a_k = arg\ max_{a_m \in A, e_{ml}=1, e_{im}=0}\ Y(a_m) \qquad (4)$$

Such strategy represents how the knowledge of the network can be used in a MAS. In this case, the task agent is making use of the centralization degree of the SNA theory. This is a measure of power, influence, and which in this case is a measure of performance. Thus, an agent is able to adapt its social network site so that this attempt to get more success in being a member of a team.

### 4.3. Communication Protocols

All SNA-based reasoning may be mapped using a set of communication protocols in order to establish a standard mechanism for establishing team for solving tasks. This protocol consists of a set of messages with ACL (Agent Communication Language) perfomatives that indicate what actions are being performed on the social network of agents.

The process begins when the manager agent sends a propose message to the scheduler agent. The scheduler agent sends this message to all task agents. If a particular task agent has one of the skills required by the task, this sends an accept-proposal message to the scheduler agent.

After receiving all the proposals, the scheduler agents reasons using SNA metrics and sends accept-proposal messages for the agents that were accepted, and reject-proposal for those that were rejected. Agents who received the acceptance, are now committed to a team and will no longer respond to the proposals made by the scheduler. The scheduler agent checks after a time $\gamma$ if the team was complete. If it is not complete, this sends a failure message to staff agents of this team, stating the fact. When a team is complete, the scheduler agent considers that the task was performed with success, sending a message confirm, releasing the agents in order to be a member of new teams.

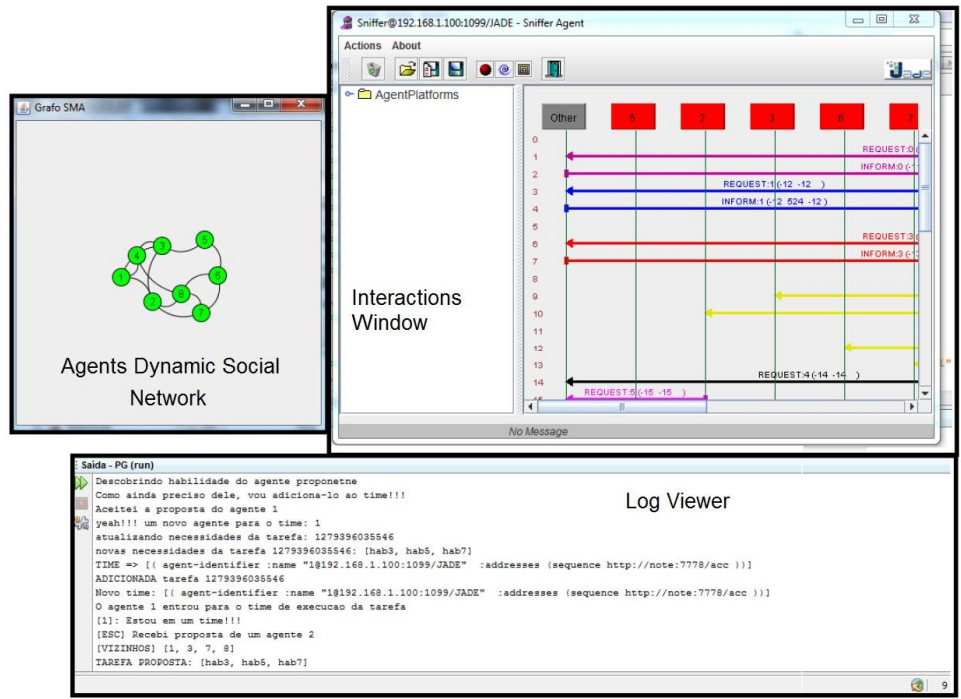

Figure 2: Prototype developed in JADE platform: social network dynamically built by agents, monitoring of messages exchanged and log of actions.

The mechanism of adaptation by performance and reference was implemented as follows: when an agent reaches a valid value of performance (for example, after 10 attempts of formation of teams), it sends a query-if message questioning to its neighbors about their values of performance. The neighbors receive this message and respond to the agent with the value of its respective performances, with a inform-if message. After receiving all the performances, the agent checks if is advantageous to perform the change. If yes, this sends a proxy message to its neighbor with higher performance by asking another agent by reference. After receiving this referenced agent the agent adds it to its network, removing the neighbor with the lowest performance.

Figure 2 shows the control screen developed in JADE platform, allowing to observe in real-time the changes in the structure of the social network, the messages exchanged between them, as well as a detailed action log, which enables us to analyzing the SNA metrics used for decision-making by each agent.

## 5. Final Considerations

Multi-agent systems and social networks are closely united. These two theories help to shape and understand social phenomena in many different ways. This work was concerned to understand how social networks could improve the multi-agent system modeling process.

The integration between the two areas can be in two ways: macro and micro modeling. The macro modeling consists in using the existing metrics in SNA to analyze multi-agent behaviors. Given the inherent distribution of multi-agents systems and the high degree of interaction in the MAS society, it is important to have a tool that assists in validation of proposed model.

On the other hand, in the micro modeling agents are built making use of social networking metrics in their behavior. With this, the interaction between the agents can generate new social phenomena, as well as allow them to take a decision based on your social network. The prototype used this approach and showed how an agent society can be built to solve a set of tasks where teams are formed based on the social network structure.

As future work, we intend to improve the modeling of new social behaviors, such as anarchy, lack of collaboration, and new strategies of relationship between agents that allow a multi-agent society to act on non-cooperative scenarios.